\documentclass[a4paper]{jpconf}
\usepackage{graphicx}
\usepackage{lineno}
\usepackage{xspace}
\usepackage{hyperref}

\newcommand{\matriplex}{\textsc{Matriplex}\xspace}
\newcommand{\cplusplus}{{\small\textsc{C++}}\xspace}

\begin{document}

\renewcommand{\pt}{\ensuremath{p_T}\xspace}

\title{Traditional Tracking with Kalman Filter on Parallel Architectures}

\author{
Giuseppe Cerati$^1$, 
Peter Elmer$^2$, 
Steven Lantz$^3$, 
Ian MacNeill$^1$, 
Kevin McDermott$^4$, 
Dan Riley$^4$, 
Matev\v{z} Tadel$^1$, 
Peter Wittich$^4$, 
Frank W\"urthwein$^1$, 
Avi Yagil$^1$
}
\address{$^1$ University of California - San Diego , La Jolla, CA, 92093, USA}
\address{$^2$ Department of Physics, Princeton University, Princeton, NJ 08540, USA}
\address{$^3$ Center for Advanced Computing, Cornell University, Ithaca NY 14853, USA}
\address{$^4$ Laboratory of Elementary Particle Physics, Cornell University, Ithaca NY 14853, USA}

\ead{Peter.Elmer@cern.ch}

\begin{abstract}
Power density constraints are limiting the performance improvements
of modern CPUs. To address this, we have seen the introduction of
lower-power, multi-core processors, but the future will be even
more exciting. In order to stay within the power density limits but
still obtain Moore's Law performance/price gains, it will be necessary
to parallelize algorithms to exploit larger numbers of lightweight
cores and specialized functions like large vector units. Example
technologies today include Intel's Xeon Phi and GPGPUs.
Track finding and fitting is one of the most computationally
challenging problems for event reconstruction in particle physics.
At the High Luminosity LHC, for example, this will be by far the
dominant problem. 
The most common track
finding techniques in use today are however those based on the
Kalman Filter. Significant experience has been accumulated with
these techniques on real tracking detector systems, both in the
trigger and offline. 
We report the results
of our investigations into the potential and limitations of these
algorithms on the new parallel hardware. 
\end{abstract}

\section{Introduction}

Track finding and fitting is one of the most computationally
challenging problems for event reconstruction in particle physics.  At
the High Luminosity LHC (HL-LHC), for example, this will be by far the
dominant problem. The need for greater parallelism has driven
investigations (e.g.\ ~\cite{Funke:2014dga, Halyo:2013gja,
  Rohr:2012nf}) of very different track finding techniques including
Cellular Automata or returning to Hough Transform techniques
originating in the days of bubble chambers. The most common track
finding techniques in use today are based on the Kalman
Filter~\cite{Fruhwirth:1987fm}. Significant experience has been
accumulated with these techniques on real tracking detector systems,
both in the trigger and offline. They are known to provide high
physics performance, are robust and are exactly those being used today
for the design of the tracking system for HL-LHC.  We report the
results of our investigations into the potential and limitations of
these algorithms on the new parallel hardware.

\begin{figure}[tbp]
  \centering
  \includegraphics[width=0.5\linewidth]{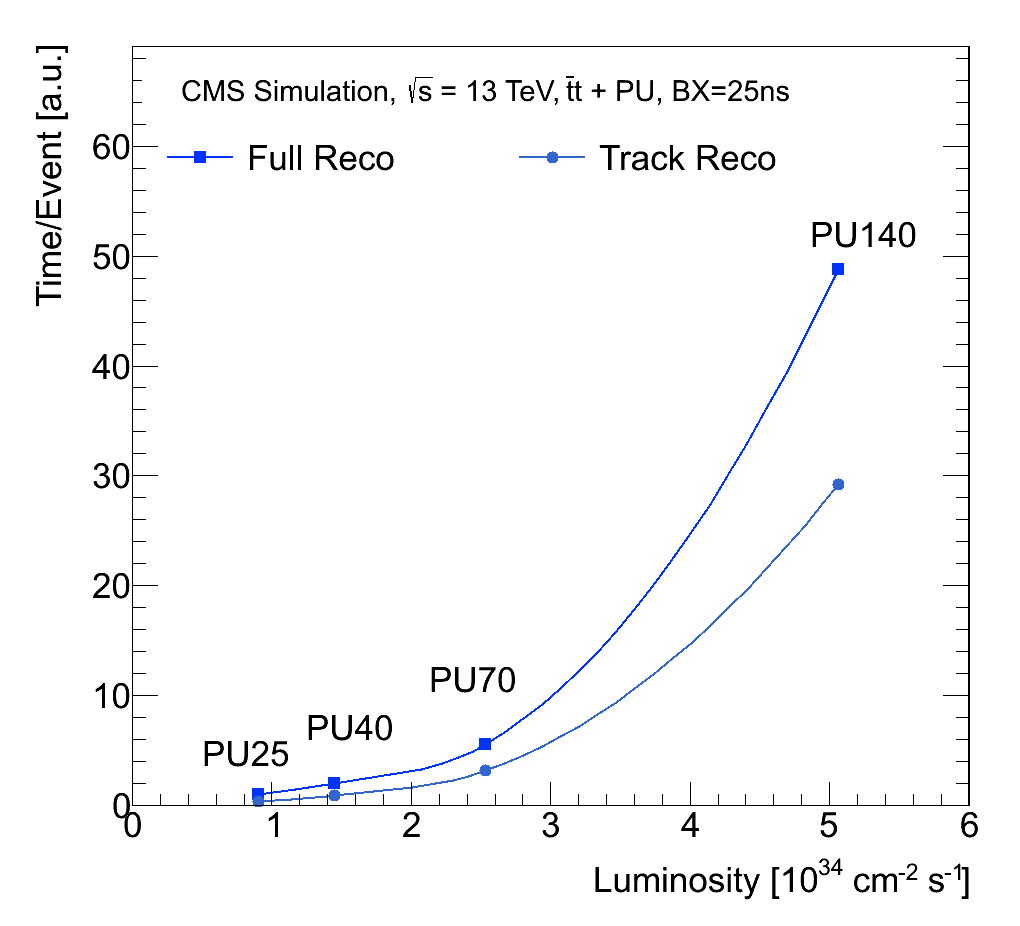}
  \caption{Timing vs instantaneous luminosity for $pp\to t\bar{t}$
    simulated events, with overlaid pile-up events, with 25 ns bunch
    crossing.  Samples with different average pile-up are used as
    reported on the plot.  We show the timing of the full CMS
    reconstruction, and of the iterative tracking sequence by
    itself. As can be seen, the time required for tracking dominates
    the total time and increases sharply at higher values of the
    instantaneous luminosity.   }
  \label{fig:cpu_tracking_pileup}
\end{figure}
{\bf Kalman Filter tracking:}
We choose to attack charged particle tracking as it is one of the most
CPU-time intensive tasks in the hadron collider event reconstruction
chain. The CPU processing time is also acutely sensitive to increases
in instantaneous luminosity and, therefore,
pileup. Figure~\ref{fig:cpu_tracking_pileup} shows the CPU time for a
typical event reconstruction job. The growth term, \emph{i.e.}, the
increase in CPU time as a function of instantaneous luminosity, is
exponential and the dominant contribution clearly comes from track
reconstruction~\cite{Vertex14}.

We choose a Kalman Filter-based algorithm as this high-performance
algorithm is the standard in modern HEP experiments. For instance, the
CMS experiment uses a Kalman Filter in both its high-level trigger as
well as its offline reconstruction
application~\cite{Chatrchyan:1704291}. The algorithm has
well-understood timing performance characteristics in serial computing
environments. The physics performance is also well understood in terms
of efficiencies and fake rates.

Additionally, the Kalman Filter algorithm is adaptable  for the wide
vector units of modern CPUs and Xeon Phi. The algorithm can  be
implemented via matrix manipulations, which are well-suited for these
hardware platforms.

To be able to harness new computing platforms, event-level
parallelization across cores is the key to gaining more speed. Due to
limitations in memory and I/O bandwidth, it is insufficient to run the
serial algorithm for many events at once in an `embarrassingly
parallel' approach.

{\bf Many-core:}
There has been a sea change in the progess of computing in the past
decade. Previously, one could expect that a given piece of code would
run progressively faster with each generation of new processors with
exponential gains consistent with Moore's-law-like growth. These gains
were largely achieved by increases in the processor clock
frequency. However, these gains have stalled in the last decade as
processors have started to hit scaling limits, as has been widely
recognized~\cite{GAMEOVER,Elmer:2013eja}. The limits are largely
driven by constraints on power consumption.

Moore's Law itself, which states that the number of transistors in an
integrated circuit doubles approximately every two years, is still
alive and kicking, but exists now in new forms. The first new form is
the advent of so-called `multi-core' processors, with more than one
functional processor on a chip, however without increases in the main
processor frequency. The next anticipated change is the advent of
so-called `small core' processors, such as ARM, GPUs, or MIC (Intel
Xeon Phi). These small cores individually have less processing power
than the big cores they replace, but in aggregate offer an immense
amount of computing power. Sequential applications, such as those
favored by HEP experiments, do not run out of the box on these
platforms, however, and must be adapted.

As an example of deploying a complex algorithm on a many-core system,
we explore the implementation of the CMS Kalman Filter tracking code
on an Intel Xeon Phi. 

\section{Challenges} 
New architectures are ``easy'' to deploy for simple problems and
tailored algorithms. This is not the problem we are facing; we need to
migrate tools from a running experiment such as CMS at the CERN
LHC. The serial algorithms currently in use have been optimized by
many person-years of effort and set a high bar for the physics
performance that we must meet before being able to deploy a many-core
system.  Existing algorithms require new thinking.  Additional
constraints come from the heterogeneous environment that we expect to
use. Software for modern HEP experiments must run on a diverse array
of platforms (basically, whatever is available across countries and
continents.)  Our code must be usable on big cores, little cores, and
GPUs; we cannot use algorithms that only work well on one
architecture.  We are exploring a family of algorithms consistent with
these constraints. Our current attempt focuses on the Intel Xeon Phi.

To exploit the Intel Xeon Phi, we need to vectorize and multithread a
Kalman Filter algorithm for track building and fitting.  Some of the
expected difficulties include the fact that the algorithm uses small
matrices, and has multiple decision points along the building of each
track.  Details of cache occupancy, pipelining, and latencies must be
understood to achieve maximum possible performance. An expected
benefit is that vectorization and multithreading should improve the
serial (big-core) performance as well, which would immediately be
applicable to a running experiment without the need for new hardware
purchases.

\section{Detector, Geometry, Simulation}

For this study, we used a simplified detector geometry consisting of
equally spaced, fixed length concentric cylinders, with radii at fixed
steps (ten layers, 2~m long, spaced every 4~cm in radius).  This was
implemented using \textsc{usolids}, a light-weight geometry library
with interfaces based on \textsc{ROOT} and \textsc{Geant4} libraries
which was created as part of the AIDA project~\cite{CERNAIDA}.  The
primary advantage of using an existing geometry library is that more
complex geometries (polygonal geometry, z-segmentation, etc.) will be
straightforward to implement as algorithm development progresses.  We
are also investigating eventual use of \textsc{VecGeom} from
\textsc{GeantV} for better vectorization.  To provide input data for
the Kalman Filter tests, a standalone simulation was written which
uses this full geometry for track propagation and simulates the
detector resolution via Gaussian smearing of the hit positions.

We generate sets of tracks with transverse momentum in the range $0.5
< \pt < 10.0$~GeV. The tracks are limited to be within the central
region of a barrel detector ($|\eta| < 1.0$) and generated uniformly
in $\phi$.  In this configuration, every track crosses every layer of
our simplified detector. 
At the intersection of the generated tracks with the detector, we
create a hit whose position is smeared in $x-y$ by
$\sigma_{xy}=0.01$~cm and in $z$ by $\sigma_z=0.1$~cm.

\section{Tracking}
\begin{figure}[tbp]
  \centering
  \includegraphics[width=0.6\linewidth]{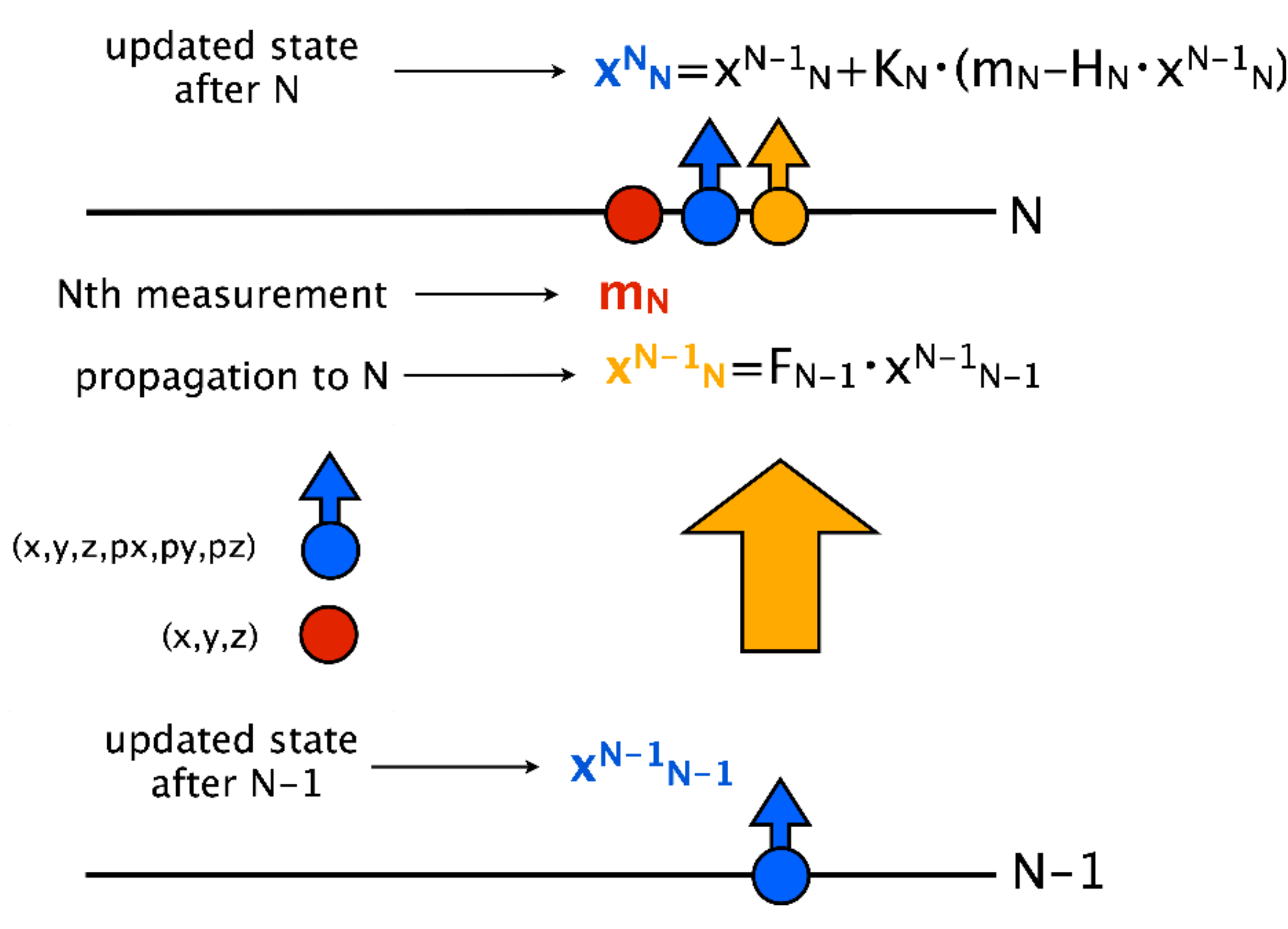}
  \includegraphics[width=0.25\linewidth]{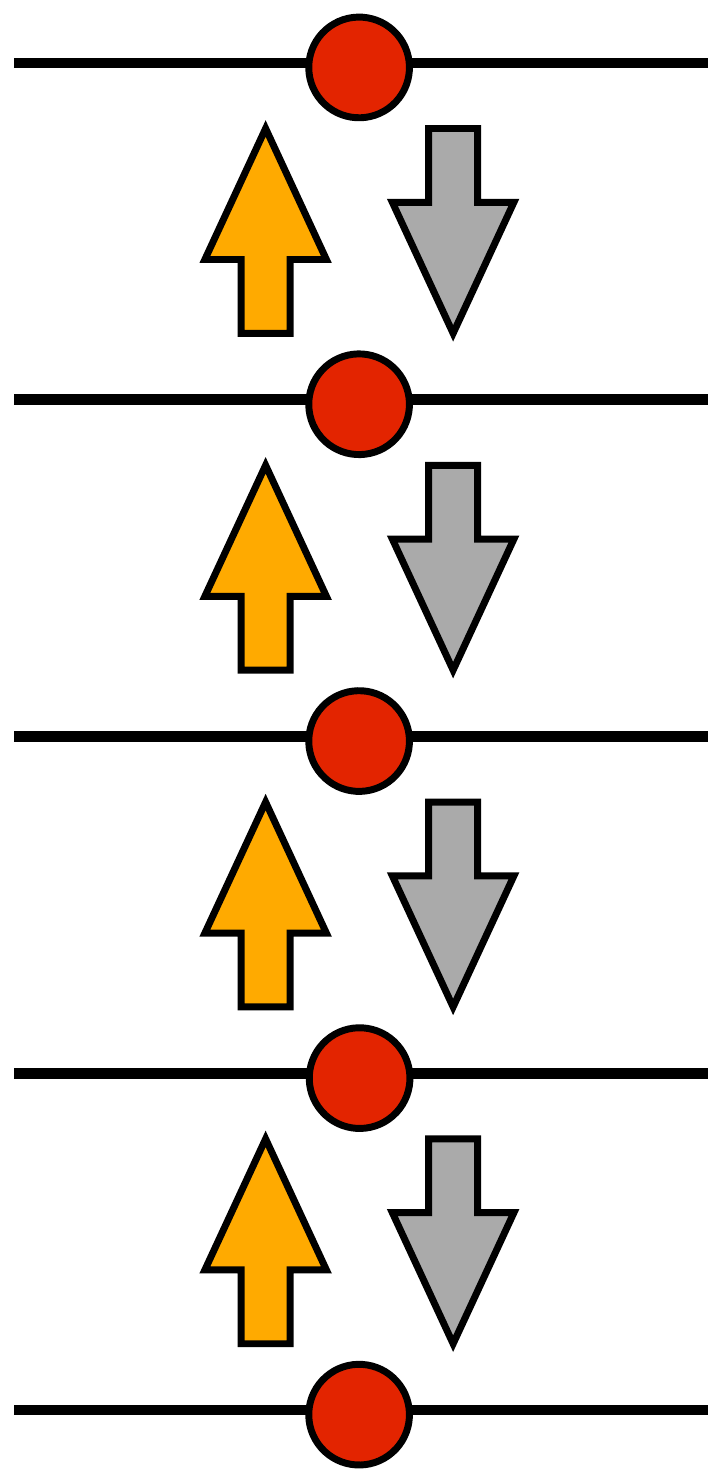}
  \caption{Left: Basic unit of the Kalman Filter algorithm. At each
    step, position information from hits is used to estimate the track
    parameters and their uncertainties.  The red circle represents the
    measurement (a hit). The blue point on layer $N$ represents the
    estimated state (position and direction) at layer $N$ before
    taking into account information from hits on that layer. The
    yellow point is the updated state at layer $N$, taking into
    account all hits from up to and including layer $N$. Right:
    Cartoon representing the two stages of fitting: forward fit and
    backward smoothing. For this test we do not perform a smoothing step. }
  \label{fig:basicunit_fitting}
\end{figure}
Track reconstruction can be divided into three steps: seeding,
building, and fitting.  While seeding is based on a different kind of
algorithm, both track building and fitting are often based on a Kalman
Filter.  The Kalman Filter is an iterative procedure where a basic
logic unit is applied repeatedly. The unit consists of the propagation
of parameters and uncertainties (\emph{track state}) from one layer to
the next. At each layer, the track state is updated with the hit
measurement information of the hits on that layer. (See
Fig.~\ref{fig:basicunit_fitting} left.) 

As can be seen in Fig.~\ref{fig:basicunit_fitting} (right), the track
fit consists of the simple repetition of the basic logic unit for all
the pre-determined track hits and therefore it is the easiest case to
test.  It is divided in two steps: a forward fit and a backward
smoothing stage for optimal performance.  For this test, only forward
fit is run.  The Kalman Filter requires initial estimation of track
parameters to get started.  In our test, the starting state is taken
directly from simulation with 100\% uncertainty.  As a more realistic
option, we also implemented a parabolic fit in the conformal space to
define initial parameters.
\begin{figure}[tbp]
  \centering
  \includegraphics[width=0.45\linewidth]{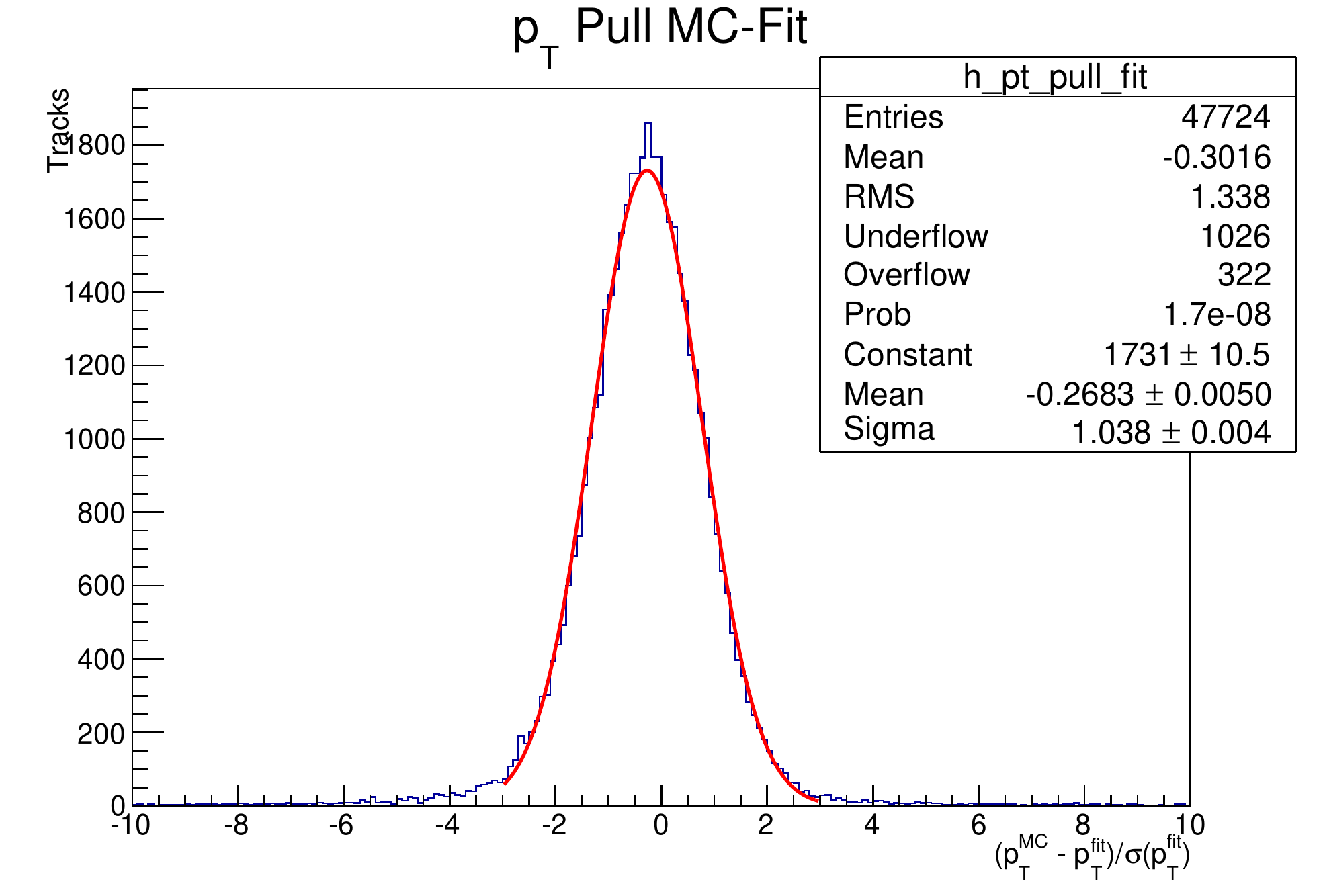}
  \includegraphics[width=0.45\linewidth]{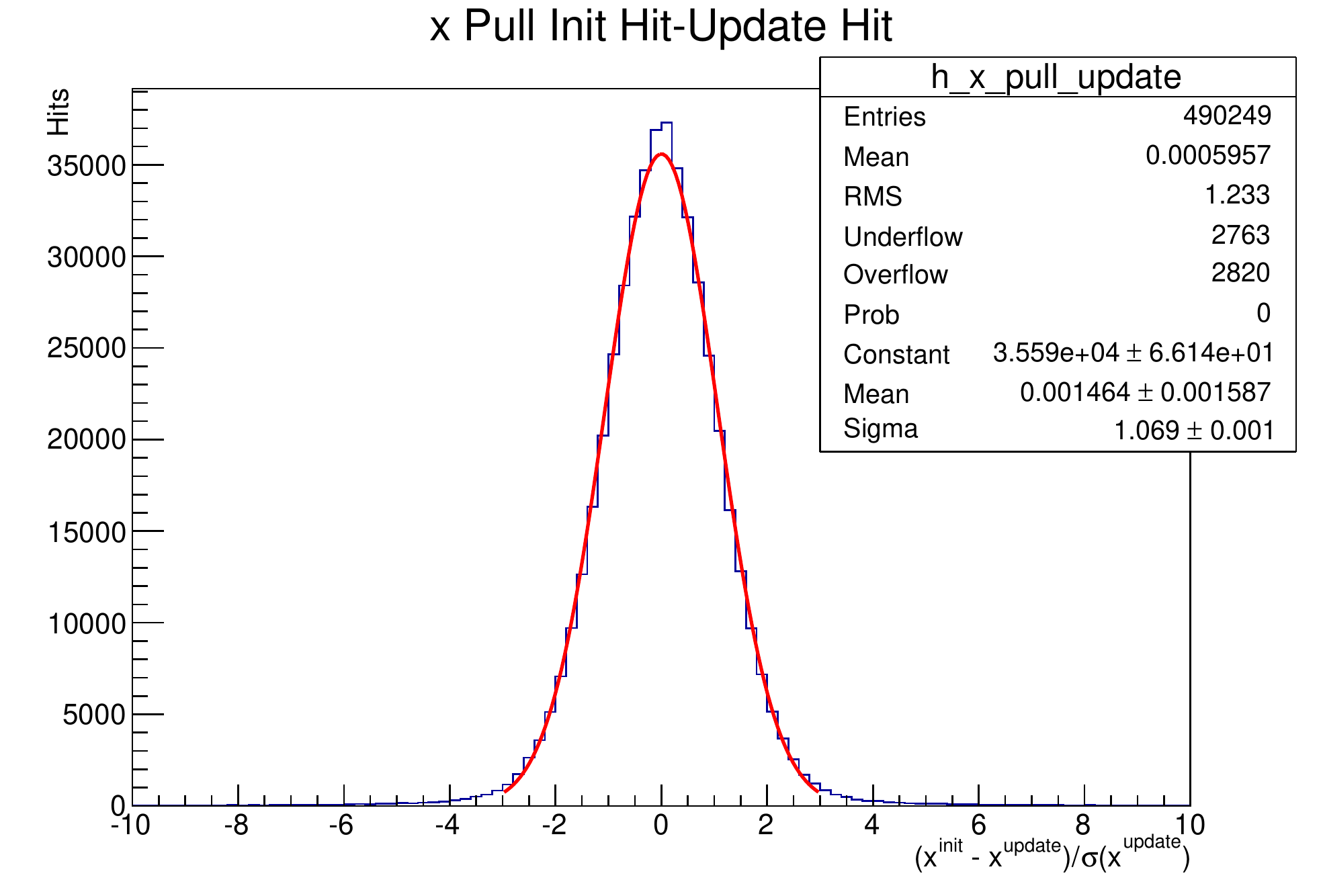}
  \caption{\pt and position pull plots for the fit. The performance of
  the fit is under control.}
  \label{fig:pulls}
\end{figure}
For the fitting
test, the hits are attached to a track ``by name'' (\emph{i.e.}, no
pattern recognition is performed) and a Kalman Filter fitting stage is
performed. Figure~\ref{fig:pulls} shows the \pt and position pulls for
the resulting fit with Gaussian distributions consistent with unit
width and demonstrates that the fit results are reasonable.
The achieved \pt resolution is roughly  $\sigma_{\pt}/\pt = 0.005\times \pt$.


\section{Optimized Matrix Library \matriplex}
\begin{figure}[tbp]
  \centering
  \includegraphics[width=0.95\linewidth]{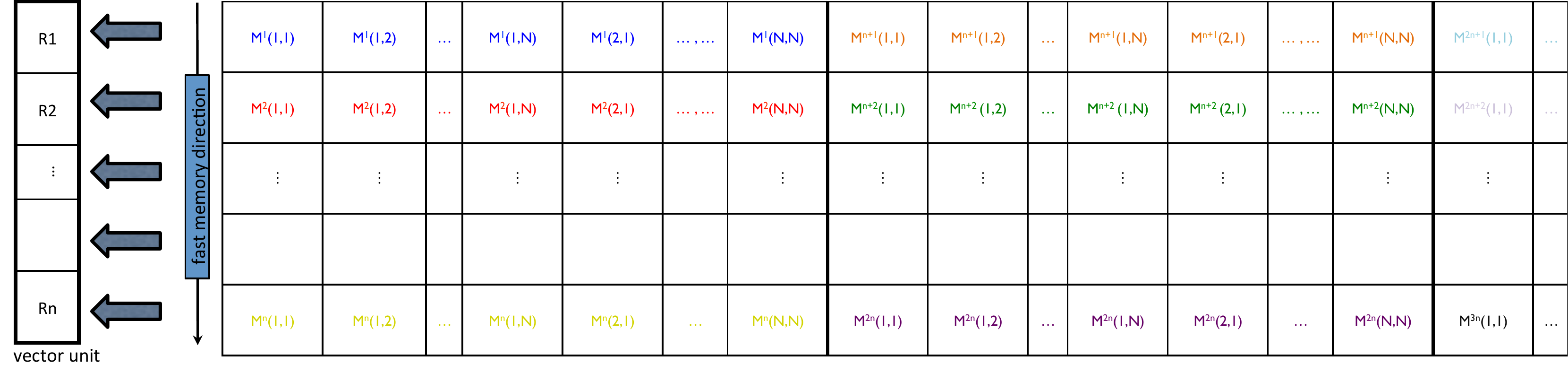}
  \caption{Memory layout for the new matrix library 
    \matriplex. The layout is optimized for our problem, which
    consists of matrix manipulations of low-dimensional matrices. The
    memory layout is matrix-major. In the Figure, the matrix dimension
  is $N\times N$ and the vector unit size is $n$.} 
  \label{fig:matriplex}
\end{figure}
The computational problem of Kalman Filter-based tracking consists of a sequence
of matrix operations on matrices of sizes from $N\times{}N = 3\times{}3$ up to 
$N\times{}N = 6\times{}6$. To allow
maximum flexibility for exploring SIMD operations on small-dimensional
matrices, and to decouple the specific computations from the high level
algorithm, we have developed a new matrix library, \matriplex. The \matriplex
memory layout is optimized for the loading of vector registers for SIMD operations
on a set of matrices as shown in Fig.~\ref{fig:matriplex}. \matriplex
includes a code generator for generation of optimized matrix operations
supporting symmetric matrices and on-the-fly matrix transposition. Patterns of
elements which are known by construction to be zero or one can be specified,
and the resulting generated code will be optimized accordingly to reduce unnecessary register
loads and arithmetic operations. The generated code can be either standard
\cplusplus or simple intrinsic macros that can be easily mapped to
architecture-specific intrinsic functions.

\section{ Results}
We present the results of this study in two stages: vectorization and
parallelization. In the first step we restructure the code to allow
use of the vector units in Xeon\footnote{CentOS 6.5, $2\times6$ core Xeon
  E5-2620 @ 2GHz, 64 GB RAM, turbo off, hyperthreading enabled} and
Xeon Phi\footnote{Xeon Phi 7150, 16 GB RAM, 61 cores @ 1.24GHz}
processors. In the second step we use \textsc{OpenMP} to parallelize
the vectorized fitting procedure across the cores on the large-core
and small-core devices.

\begin{figure}[tbp]
  \centering
  \includegraphics[width=0.45\linewidth]{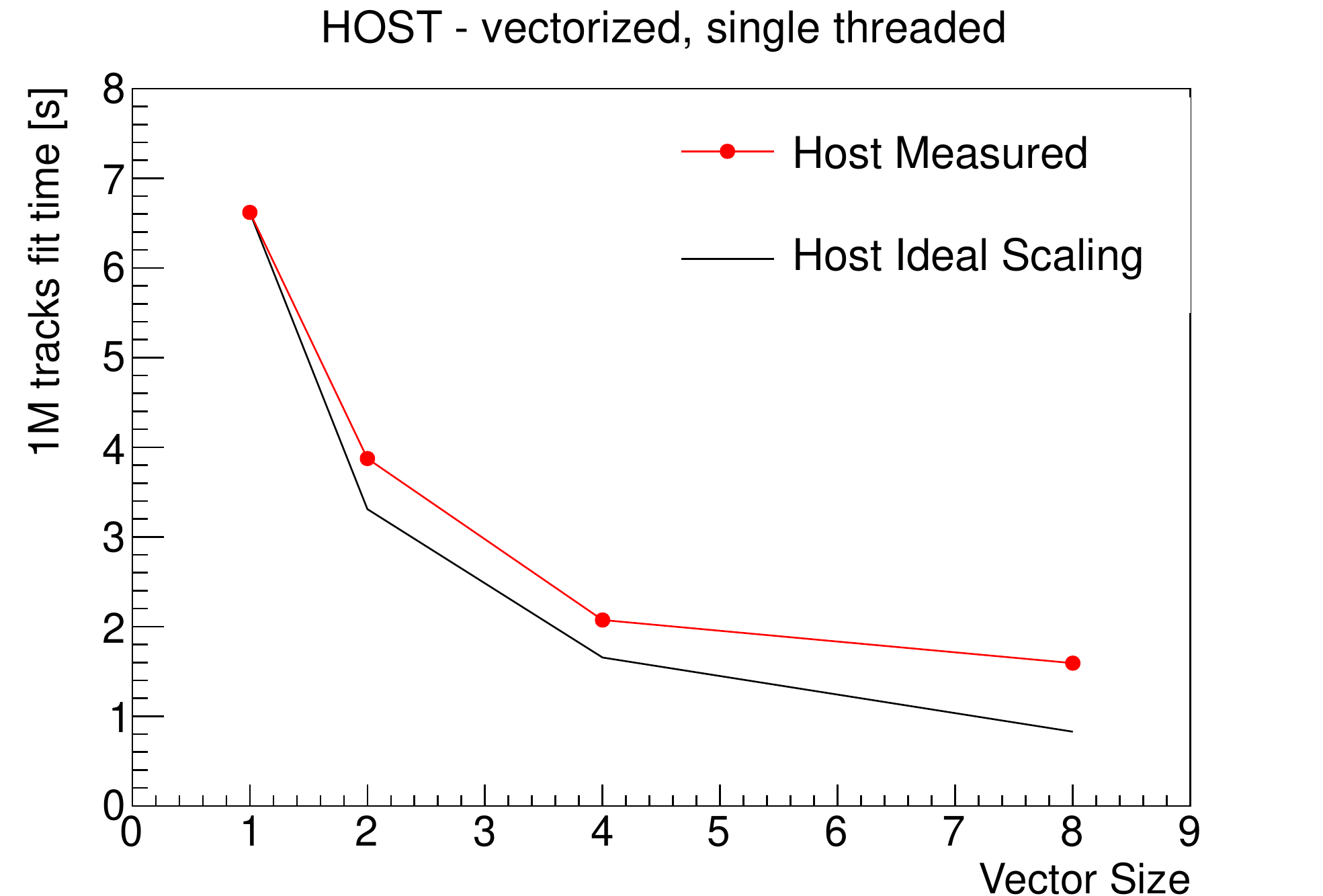}
  \includegraphics[width=0.45\linewidth]{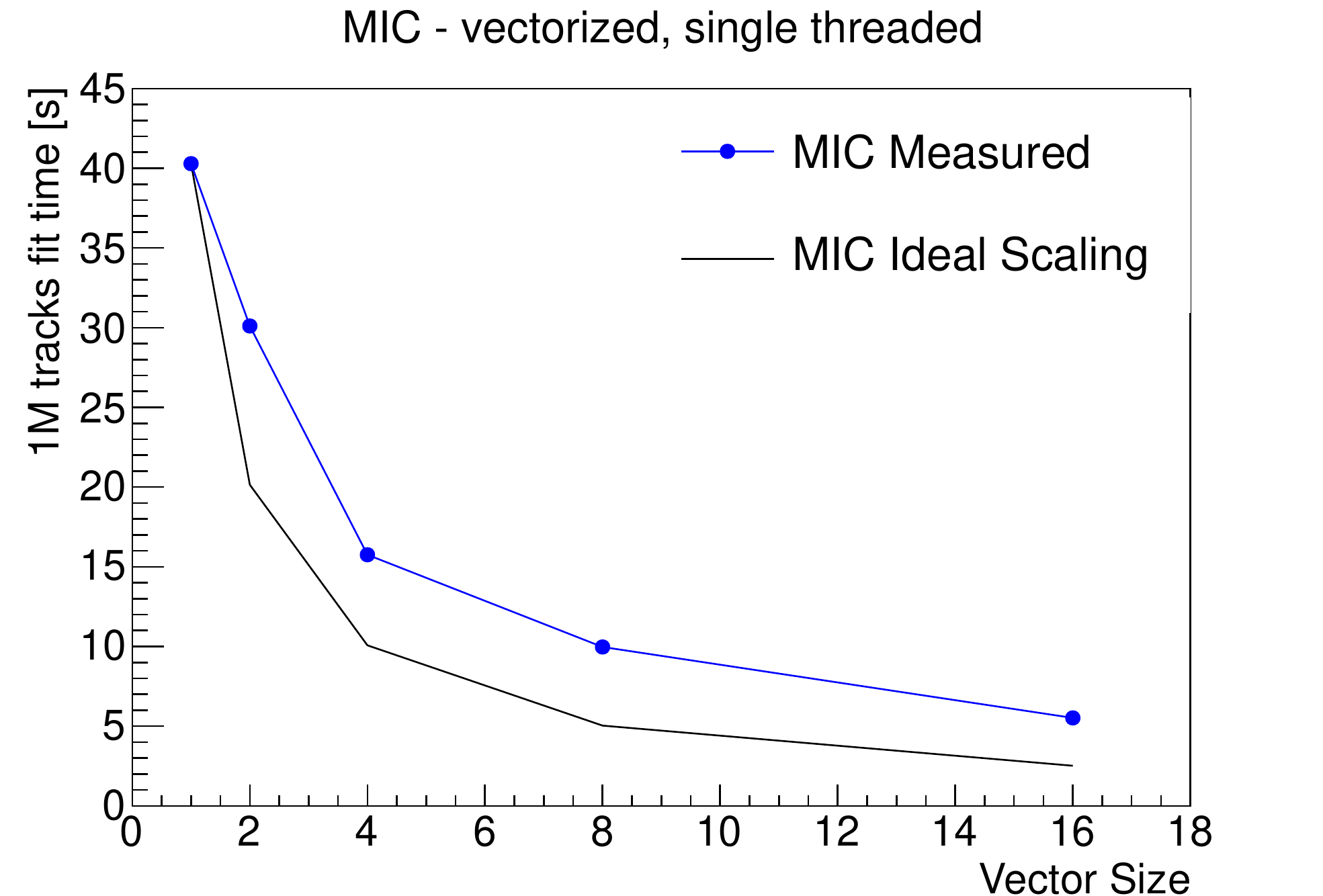}
  \caption{Timing results for vectorized code, as a function of the
    vector size, for host (left) and MIC (right). The results are
    compared to an ideal scaling described in the text. Significant
    speedups compared to serial code are observed. }
  \label{fig:vectorization}
\end{figure}

Figure~\ref{fig:vectorization} shows the timing for fitting 1M tracks
as a function of the vector size, using a single thread.  Results are
compared to scaling of serial processing time (``ideal scaling''),
defined as the time with vector unit size=1 divided by the vector unit
size.  Both for Xeon and Xeon Phi, a significant vectorization speedup
is achieved, with an effective utilization of the vector units of
$\sim50\%$ .

\begin{figure}[tbp]
  \centering
  \includegraphics[width=0.45\linewidth]{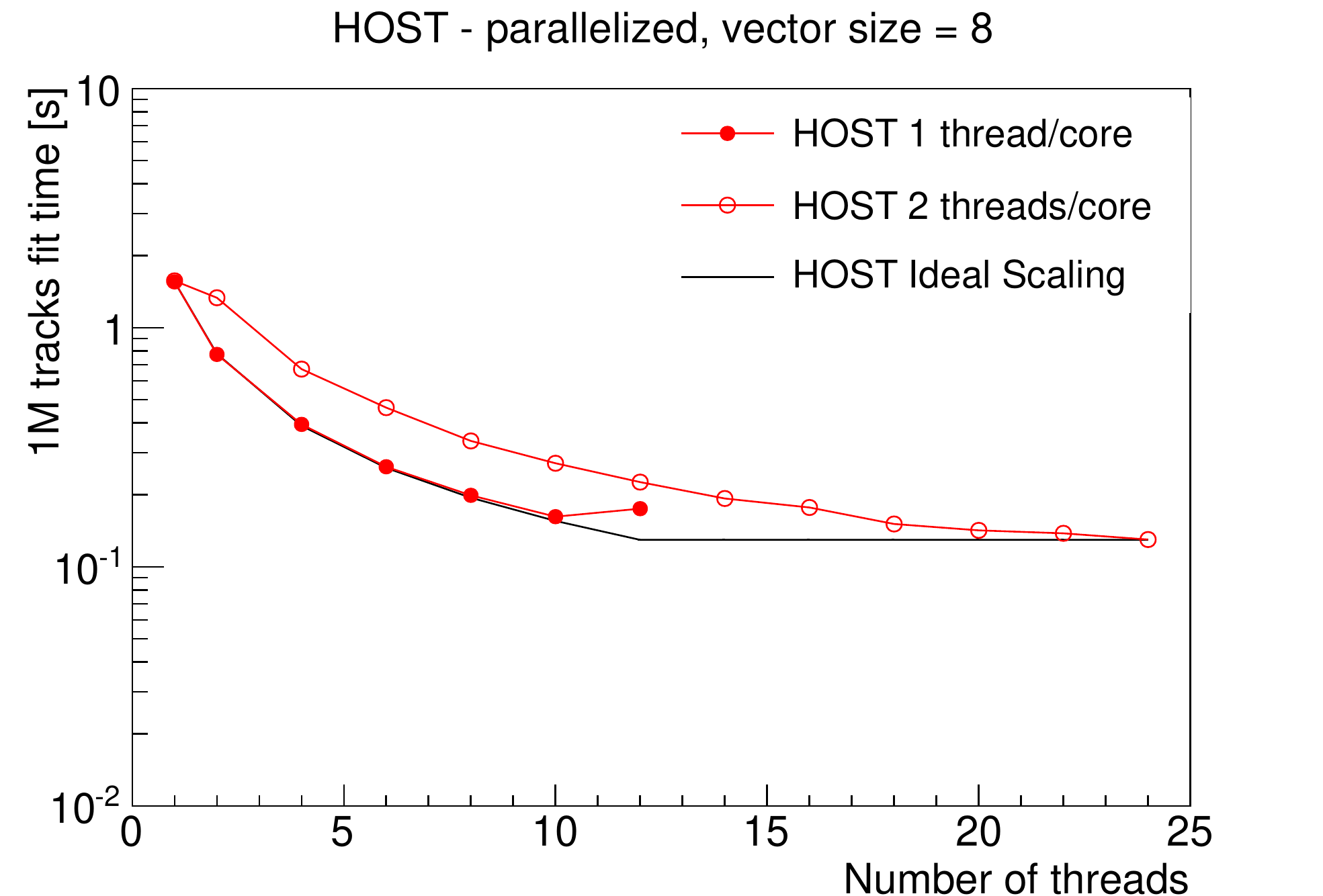}
  \includegraphics[width=0.45\linewidth]{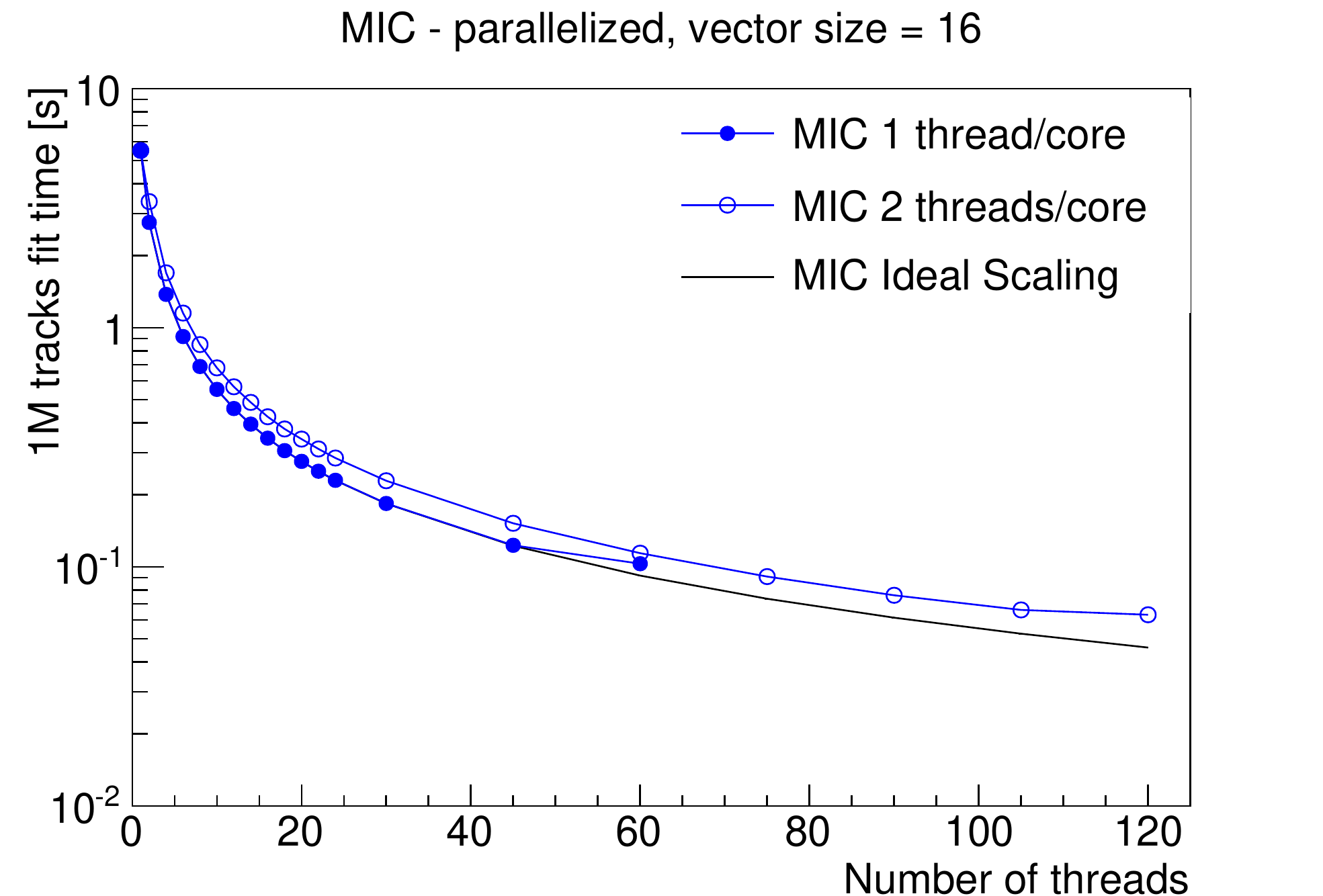}
  \caption{Timing results for parallelization tests, as a function of
    the number of threads, for host (left) and MIC (right). Two
    different methods of distributing threads across cores is shown,
    and compared to an ideal scaling. We observe ideal scaling when
    distributing one thread/core.}
  \label{fig:parallel}
\end{figure}

Figure~\ref{fig:parallel} shows the timing for fitting the same set of
tracks as a function of the number of threads, in case all vector
units are used.  We test two approaches for distributing threads on
the cores: filling every core with one thread or adding a second
thread on the same core before moving to a different one.  We compare
to ideal parallelization performance (``ideal scaling''), assuming no
hyperthreading (\emph{i.e.} a maximum of 12 threads on Xeon, maximum
120 threads on Xeon Phi).  Performance for one thread/core approach
follows the ideal curve, with a small overhead only when all cores are
filled.  The two threads per core approach shows deviation from ideal
behavior immediately: on Xeon, this is due to the use of
hyperthreading slots, while on Xeon Phi we surmise it is related to L1
cache contention issues.

\section{Future Work: Track Building}

\begin{figure}[tbp]
  \centering
  \includegraphics[width=0.25\linewidth]{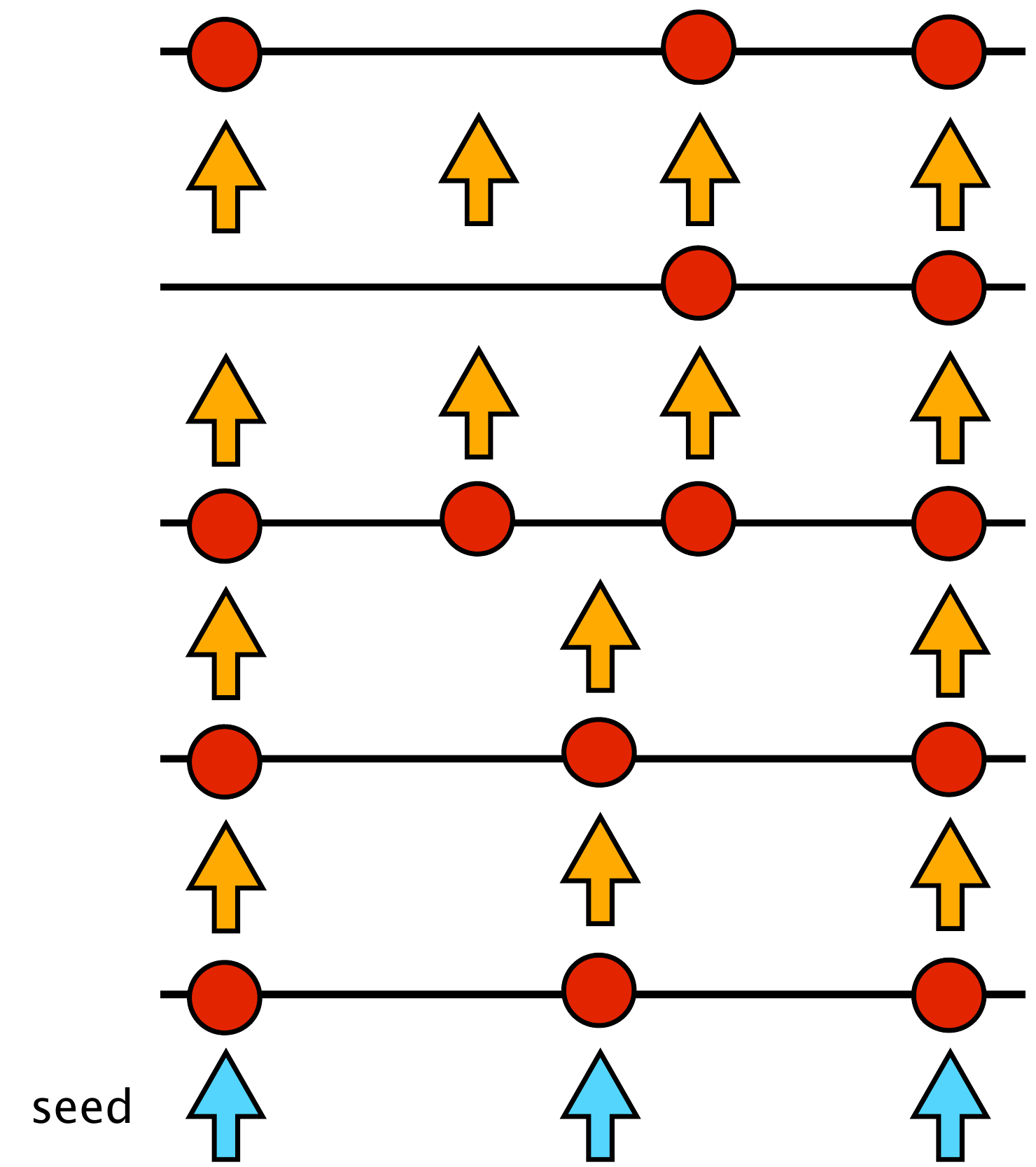}
  \caption{Schematic representation of track building. Unlike in track
    fitting, the algorithm has many branch points, \emph{e.g.} when
    missing hits on layers or when multiple hit candidates are
    encountered on a layer. This is among the challenges for achieving
    efficient vectorization of the track building code. }
  \label{fig:trackbuild}
\end{figure}
Track building is by far the most time-consuming step in the whole
event reconstruction and thus it is the most important next step of
development.  With respect to track fitting, track building adds
significantly more complexity to the problem as hits are
searched for within a compatibility window when moving to the next layer.
The track candidate needs
to branch in case of multiple matches, and the algorithm needs to be
robust against missing (due to detector inefficiency) or outlier hits
(Fig.~\ref{fig:trackbuild}).  We implemented a preliminary serial
version and tested it on events with 500 simulated tracks/event,
finding a hit finding efficiency close to 100\%.  The inherent
branching of the algorithm, along with the variable size of the
combinatorial problem, make it difficult to vectorize and parallelize
it efficiently; thus, specific design choices have to be made to boost
its computing performance on the coprocessor.

\section{Progress and Plans}

To summarize, we presented the start of a task of adapting a complex
algorithm from running HEP experiments for use in new parallel
hardware. We showed a preliminary implementation of Kalman
Filter-based track fitting, vectorized and parallelized on Intel Xeon and
Intel Xeon Phi.  In both cases, we achieve $\sim50$\% vectorization
efficiency and $\sim100$\% parallelization efficiency when running one
thread per core. 


We are currently working towards the implementation of more realistic
geometries, particularly polygonal/planar geometries typical of pixel
detectors, and of material effects, both in the simulation and
reconstruction models, to complete the study of track fitting. The
next major step is the implementation of a fully vectorized and
parallelized track building algorithm.  At the same time, we will
continue work on optimizing fitting and matrix operations so that
overheads are further reduced and computing efficiency is increased.

At this point the results shown encourage us to continue pursuing
implementation of the Kalman Filter algorithm for charged particle
tracking on parallel hardware, and demonstrate successful first steps
towards our goal of exploiting parallel hardware for HEP experiments.

\section*{Acknowledgements}
This work was partially supported by the National Science Foundation,
partially under Cooperative Agreement PHY-1120138, and by the
U.S. Department of Energy.  

\section*{References}
\bibliographystyle{unsrt} 
\bibliography{acat2014-tracking}

\end{document}